\documentclass[iop,apjl,revtex4,natbib]{emulateapj}
\usepackage{graphicx}
\usepackage{times}
\usepackage{amsmath}

\def\drm{{\rm d}}

\def\Rh{R_{\rm h}}

\def\slosT{\langle\sigma^2_{\rm los}\rangle}
\def\stotT{\langle\sigma^2_{\rm tot}\rangle}
\def\C{C_x}
\def\percent{\text{ per cent}}

\def\apj{ApJ}
\def\apjl{ApJL}
\def\apjs{ApJS}
\def\mnras{MNRAS}
\def\aap{A \& A}

\usepackage{color}
\definecolor{darkred}{rgb}{0.55, 0.0, 0.0}
\definecolor{darkblue}{rgb}{0.0, 0.0, 0.55}
 \definecolor{darkgreen}{rgb}{0.0, 0.2, 0.13}
\usepackage[colorlinks=true,linkcolor=darkred,citecolor=darkblue,urlcolor=darkgreen]{hyperref}

\begin{document}

\title[Mass estimators for flattened galaxies]{Mass estimators for flattened dispersion-supported galaxies}
\author{Jason L. Sanders}
\author{N. Wyn Evans}
\affil{Institute of Astronomy, Madingley Road, Cambridge, CB3 0HA}
\email{jls,nwe@ast.cam.ac.uk}
\date{Accepted XXX. Received YYY; in original form ZZZ}

\label{firstpage}
\begin{abstract}
We investigate the reliability of mass estimators based on the
observable velocity dispersion and half-light radius $R_\mathrm{h}$
for dispersion-supported galaxies.  We show how to extend them to
flattened systems and provide simple formulae for the mass within an
ellipsoid under the assumption the dark matter density and the stellar
density are stratified on the same self-similar ellipsoids. We demonstrate
explicitly that the spherical mass
estimators~\citep{Walker2009,Wolf2010} give accurate values for the
mass within the half-light ellipsoid, provided $R_\mathrm{h}$ is
replaced by its `circularized' analogue
$R_\mathrm{h}\sqrt{1-\epsilon}$. We provide a mathematical
justification for this surprisingly simple and effective workaround.
It means, for example, that the mass-to-light ratios are valid not
just when the light and dark matter are spherically distributed, but
also when they are flattened on ellipsoids of the same constant shape.
\end{abstract}

\keywords{
galaxies: dwarf --- galaxies: kinematics and dynamics --- dark matter}
\section{Introduction}

Accurate estimates of the dark matter content of dwarf spheroidal
galaxies (dSphs) are crucial for furthering our understanding of
galaxy formation and structure. Calculating reliable mass estimates
has historically been an awkward problem as with only line-of-sight
(l.o.s.) velocity measurements the mass profile of a spherical galaxy
can only be inferred by making an assumption about the degree of
velocity anisotropy i.e. the ratio of radial to tangential motion.

Through comparisons to solutions of the Jeans equations, it has been
shown that the mass contained near the half-light radius of a
dispersion-supported galaxy is approximately independent of the
velocity anisotropy and the radial profile of the dark and luminous
matter and is simply related to the half-light radius $\Rh$
and the luminosity-averaged l.o.s. velocity dispersion
$\sqrt{\slosT}$. There exist several different forms for these
formulae in the literature
\citep{Walker2009,Wolf2010,Amorisco2012,Campbell2016} that may be
summarised as
\begin{equation}
M_\mathrm{sph}(<r_x) = \frac{C_x \slosT\Rh}{G}
\label{Eqn::WalkerWolf}
\end{equation}
where $M_\mathrm{sph}(<r_x)$ is the mass contained within a sphere of radius $r_x$
and $G$ the familiar gravitational constant. $C_x$ is a constant that
depends on the choice of radius $r_x$. \cite{Walker2009} proposed that
if $r_x=\Rh$ then $C_x=2.5$ based on a simple example of the stellar
distribution following a Plummer profile and the dark matter following
a cored isothermal profile although this was validated through fuller
testing. \cite{Wolf2010} demonstrated that for $r_x\approx\tfrac{4}{3}\Rh$
(approximately the 3D spherical half-light radius for a range of
observationally-motivated profiles) that $C_x=4$ reproduced the
results from full Jeans analyses and was also shown to be
mathematically true under the assumption of a near-flat velocity
dispersion profile.

Although spherical mass estimators have proved useful for
understanding dSphs, they cannot give the full picture as they do not
consider the fundamentally aspherical shape of these galaxies.  Our
aim in this Letter is to find mass estimators equivalent to
equation~\eqref{Eqn::WalkerWolf} applicable to flattened systems.  We
begin by inspecting the validity of the spherical mass estimators and
go on to investigate the applicability of the estimator when
considering flattened systems in which the dark and light matter are stratified on the same self-similar ellipsoids. We give formulae similar to
equation~\eqref{Eqn::WalkerWolf} that may be used when the 3D shape of
the system is known. By marginalizing over prior assumptions on the
intrinsic shape and alignment, we show how the mass can be estimated
when the intrinsic shape and alignment are not known.

\section{Spherical mass estimators}
For a spherical stellar luminosity density
$j_\star(r)$ with a constant mass-to-light ratio in a spherical mass density $\rho_\mathrm{DM}(r)$ with mass profile $M(r)$ sourcing potential $\Phi(r)$, the potential energy can be written in
terms of the surface brightness $S(R)$ as
\begin{equation}
W=\tfrac{1}{2}\int\drm V\,j_\star(r)\Phi(r)=4\pi G\int_0^\infty\drm r\,I(r)M'(r),
\end{equation}
where
\begin{equation}
I(r)=\int_r^\infty\drm r\,rj_\star(r)=-\frac{1}{\pi}\int_r^\infty\drm R\,(R^2-r^2)^{1/2}{\frac{\drm S}{\drm R}}
\end{equation}
From the virial theorem, we know that the l.o.s. velocity dispersion is related to the total luminosity $L$ by $\slosT = -W/3L$ which gives the
constant $\C$ as
\begin{equation}
C_x=\frac{1}{R_\mathrm{h}}\Big[\int_0^{r_x}\drm r\,r^2\rho_\mathrm{DM}(r)\Big]\Big[\int_0^{\infty}\drm r\,r^2J(r)\rho_\mathrm{DM}(r)\Big]^{-1}
\label{Eqn::WynC}
\end{equation}
where $J(r)=(4\pi/3L)I(r)$. The constant $\C$ depends only on the profile of the halo model $\rho_\mathrm{DM}$ and
the surface brightness profile $J(r)$.

\begin{figure}
$$\includegraphics[width=0.9\columnwidth]{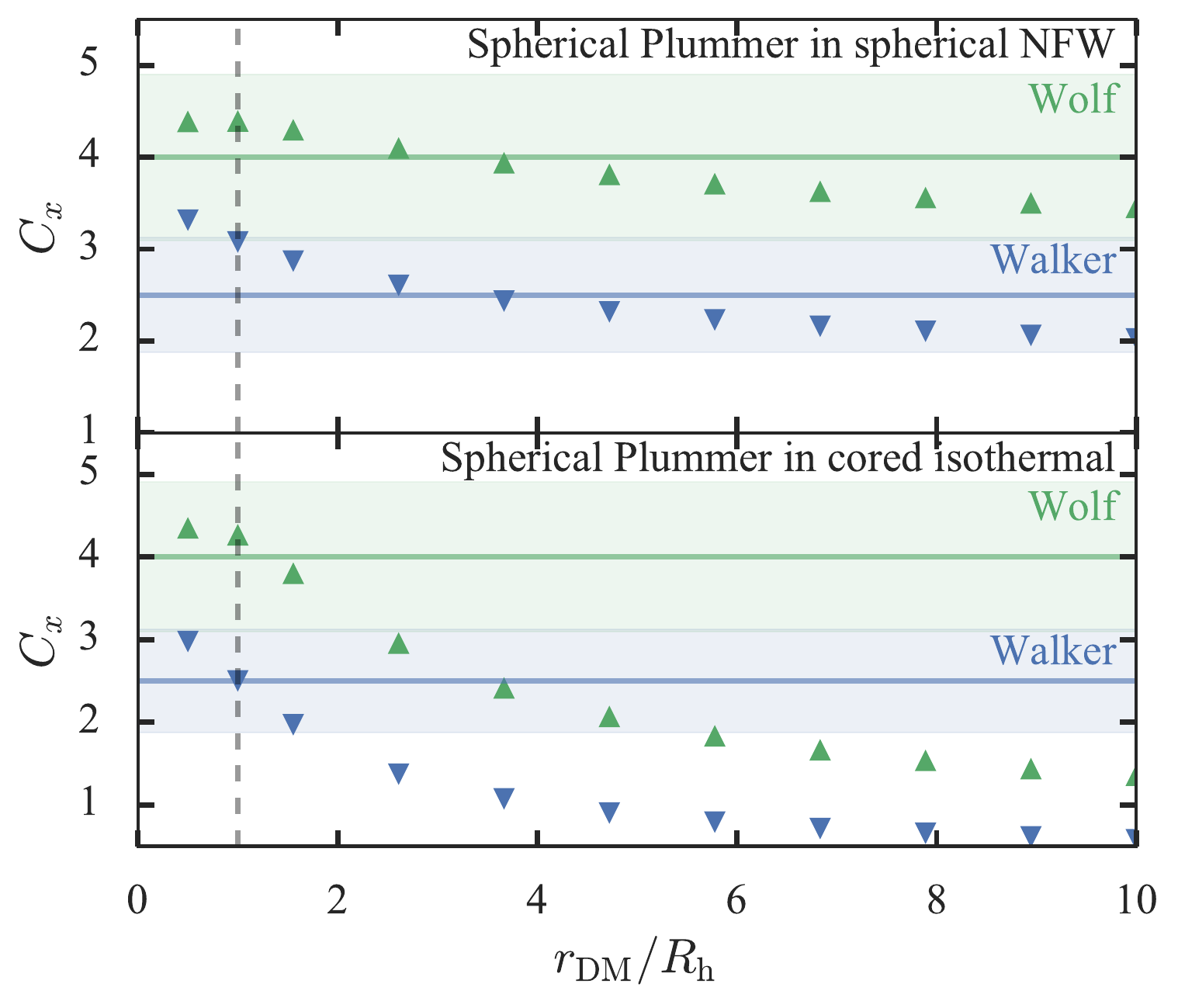}$$
\caption{Constant $\C$ in the spherical mass estimator formula against
  the ratio of the dark-matter scale radius to the stellar half-light
  radius for a Plummer model embedded in an NFW (upper panel) and cored
  isothermal (lower) halo. The blue down-pointing
  triangles show $\C$ at the radius $r=R_\mathrm{h}$, for which the
  \protect\cite{Walker2009} advocate a value of $\C=2.5$ (shown with a
  solid blue horizontal line). The green up-pointing triangles show
  the $\C$ at the radius $r=\tfrac{4}{3}R_\mathrm{h}$, for which
  \protect\cite{Wolf2010} advocate a value of $\C=4$ (green line). The
  bands show the uncertainties from \protect\cite{Campbell2016}.}
\label{Fig::SphericalMassEstimator}
\end{figure}

\begin{figure*}
$$\includegraphics[width=\textwidth]{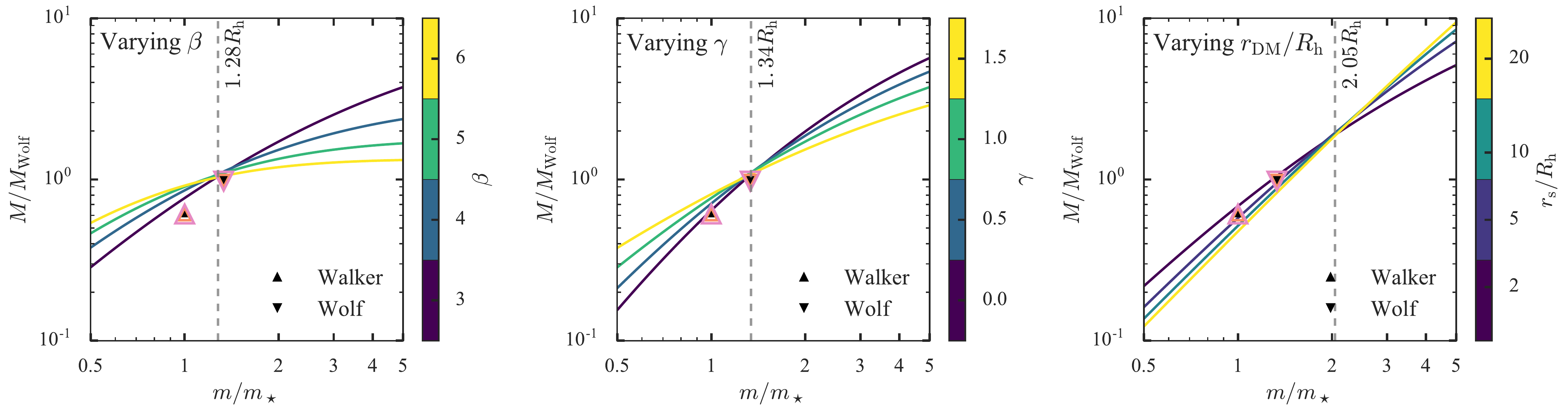}$$
\caption{Spheroidal mass profile for a stellar Plummer profile embedded in
  a double power-law dark-matter halo with varying outer
  slope $\beta$, inner-slope $\gamma$ and scale radius
  $r_\mathrm{DM}$. All models have the same luminosity-averaged
  l.o.s. velocity dispersion and half-light radius $\Rh$. The masses
  are normalized with respect to the Wolf mass estimate. The default
  parameters are $\gamma=1$, $\beta=3$ (NFW) and
  $r_s/R_\mathrm{h}=1$. The black points show the results of two mass
  estimators and the vertical dashed line shows the point of minimum
  variance in the logarithm of the mass for each set of curves. The
  spheroidal mass estimates using the mass estimator proposed in this
  paper are given for an edge-on oblate ($q=0.6$, orange triangles)
  and edge-on prolate model ($p=q=0.6$, pink triangles).}
\label{Fig::PlummerNFW_rrat1}
\end{figure*}

We use this to test the validity of the spherical mass
estimator. In Fig.~\ref{Fig::SphericalMassEstimator}, we show the
result of equation~\eqref{Eqn::WynC} computed numerically for two
models with differing ratios of dark to stellar scale-lengths
($r_\mathrm{DM}/\Rh$). They are an NFW dark matter
profile $\rho_\mathrm{DM}(r)\propto r^{-1}(1+r/r_\mathrm{DM})^{-2}$
and a cored isothermal profile of the form
\begin{equation}
\rho_\mathrm{DM}(r)=\frac{v_0^2}{4\pi G}\frac{3r_\mathrm{DM}^2+r^2}{(r_\mathrm{DM}^2+r^2)^2}.
\end{equation}
The stellar tracer profile follows a Plummer law $\rho_\star(r)\propto
(1+(r/r_\star)^2)^{-5/2}$ for which $R_\mathrm{h}=r_\star$. The
constant $\C$ is computed at the two radii recommended by
\cite{Walker2009} and \cite{Wolf2010}. The constants given by these
two authors are shown with horizontal lines along with the uncertainty
found by \cite{Campbell2016} from inspecting cosmological
hydrodynamical simulations. The variation of $\C$ with respect to
$r_\mathrm{DM}/\Rh$ is smallest for the NFW profile and is consistent
with the bracket found by \cite{Campbell2016}. In the cored isothermal
profile with $r_\mathrm{DM}/\Rh\approx1$, both estimators perform
well. However, again as $r_\mathrm{DM}/\Rh$ is increased, $\C$ deviates
significantly and so the estimators perform poorly for
$r_\mathrm{DM}/\Rh>2$.

We now explore how the mass estimators perform as the parameters of
a double power-law dark matter density profile are altered. We use a fixed Plummer
profile for the stars with a $\mathrm{sech}$ truncation at
$10R_\mathrm{h}$. In Fig.~\ref{Fig::PlummerNFW_rrat1}, we show the
mass profiles of different dark matter profiles that all produce the
same luminosity-averaged l.o.s. velocity dispersion. The default
parameters are those of an NFW profile with $r_\mathrm{DM}/\Rh=1$ and
a $\mathrm{sech}$ truncation at $10r_\mathrm{DM}$. We alter the outer
slope $\beta$, inner slope $\gamma$ and the ratio
$r_\mathrm{DM}/\Rh$. We find that when varying the inner and outer
slopes the pinch point where the mass is the same for all profiles is
around $\tfrac{4}{3}\Rh$ i.e. the radius recommended by
\cite{Wolf2010}. Varying $r_\mathrm{DM}/\Rh$ produces a pinch point
further out. This helps explain why mass estimators derived for use on
realistic halos with $r_\mathrm{DM}>\Rh$ can constrain the mass at larger
radii \citep[e.g.][]{Amorisco2012,Campbell2016}.

\section{Flattened mass estimators}

We now turn to adapting the spherical mass estimators for application
to flattened systems. We work with models with both the dark and stellar density stratified on the same
concentric self-similar ellipsoids labelled with the coordinate $m$ such that
$m^2=x^2/a^2+y^2/b^2+z^2/c^2$
with $a>b>c$. The axis ratios of the ellipsoids are $p=b/a$ and
$q=c/a$. We view the model along the spherical polar unit vector
defined by the angles $(\vartheta,\varphi)$, where $\vartheta$ is the
co-latitudinal angle and $\varphi$ the azimuthal angle defined with
respect to a Cartesian coordinate system aligned with the principal
axes (see Fig.~\ref{Fig::Diagram}. When oblate and prolate spheroids are viewed `face-on', they
appear round. The spherical mass estimator underestimates
(overestimates) the mass within a sphere for the oblate (prolate)
case, as mass is added to (removed from) the sphere. Similarly,
the formulae give (smaller) under- and overestimates for the mass
within the corresponding ellipsoid. We seek an appropriate
modification to equation~\eqref{Eqn::WalkerWolf} that is applicable to
flattened systems, namely
\begin{equation}
M_\mathrm{ell}(<m_x)=\frac{C_xf_\sigma\slosT f_r\Rh}{G},
\label{Eqn::WalkerWolf_Flattened}
\end{equation}
where $M_\mathrm{ell}(<m_x)$ is the mass within an ellipsoid (that is
the same shape as the equidensity contours) with major axis length
$r_x$. We imagine creating an ellipsoidal model by deforming a
spherical model that obeys the spherical mass estimator formulae
outlined in the previous section. The total mass is conserved if
$abc=1$ and the mass within an ellipsoid of major-axis length $r_x$ is
identical to the mass within a sphere of radius
$m_x=r_x/a=r_x(pq)^{1/3}$. However, to estimate this parent spherical
model mass from the spherical mass estimator formulae, we must relate
the observed l.o.s velocity dispersion to the spherical velocity
dispersion and the observed half-light major-axis length to the
intrinsic major-axis length of the considered ellipsoid. Assuming the
total velocity dispersion (the average of the dispersions along the principal axes) is conserved as we deform the model\footnote{To leading order in the flattening, the ratio of the total
dispersion of the flattened model to the spherical model with the same
mass is $\langle\sigma_\mathrm{tot}^2\rangle_\mathrm{flat}/
\langle\sigma_\mathrm{tot}^2\rangle_\mathrm{sph}\approx1-
\tfrac{4}{45}\Big[(1-p)^2-(1-p)(1-q)+(1-q)^2\Big]$
.},
the factor $f_\sigma$ accounts for the
relationship between the l.o.s velocity dispersion and the total
dispersion of the ellipsoidal model. The factor $f_r$ accounts for the
relationship between the observed major-axis length and the intrinsic
major-axis length of the equivalent ellipsoid (and that of the parent
spherical model).

\subsection{Velocity scaling}
For triaxial systems, the velocity scaling $f_\sigma=\stotT/\slosT$ is given by

\begin{equation}
f_\sigma=\frac{1}{3}\frac{1+r_{xz}+r_{yz}}{\cos^2\vartheta+r_{xz}\sin^2\vartheta\cos^2\varphi+r_{yz}\sin^2\vartheta\sin^2\varphi}
\end{equation}
where
\begin{equation}
r_{ij}=\langle\sigma_i^2\rangle/\langle \sigma_j^2\rangle=W_{ii}/W_{jj}\text{ (no sum)}.
\end{equation}
For dSphs in which the stellar and dark-matter density profiles are
stratified on the same self-similar ellipsoids, $r_{ij}$ depends only on the
shape of the ellipsoids \citep{Roberts1962, BinneyTremaine}. That is to say,
it is independent of the `radial' density profile of the light and dark matter.
Therefore, $f_\sigma$ is a function of $p$, $q$ and the viewing angles:
$f_\sigma=f_\sigma(\vartheta,\varphi,p,q)$. Expressions for $W_{ij}$ are
given in Table 2.2 of \cite{BinneyTremaine}.

\subsection{Radial scaling}

\begin{figure}
$$\includegraphics[width=\columnwidth]{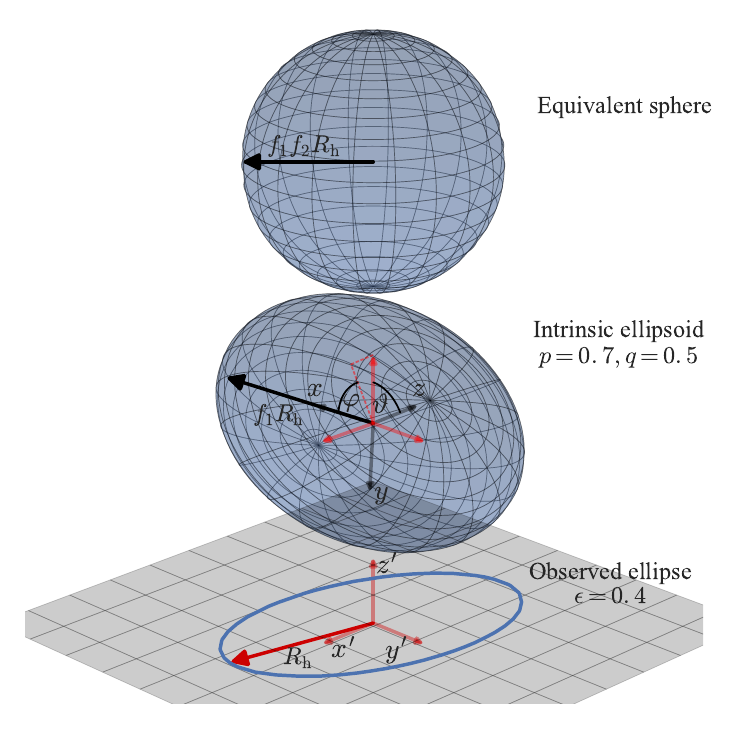}$$

\mbox{$$\includegraphics[width=.49\columnwidth]{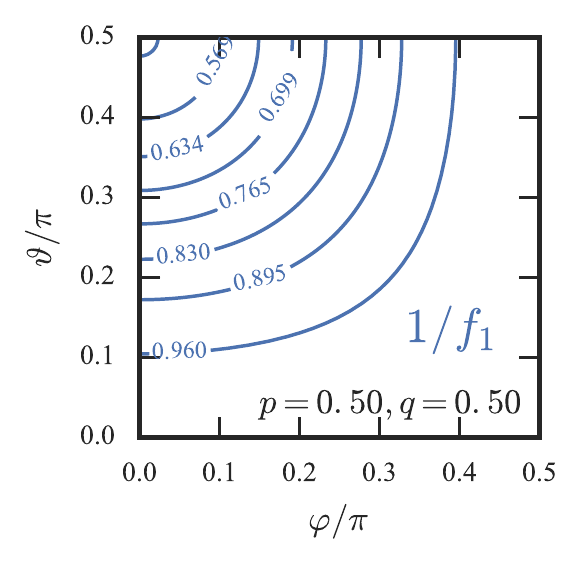}$$}
\mbox{$$\includegraphics[width=.49\columnwidth]{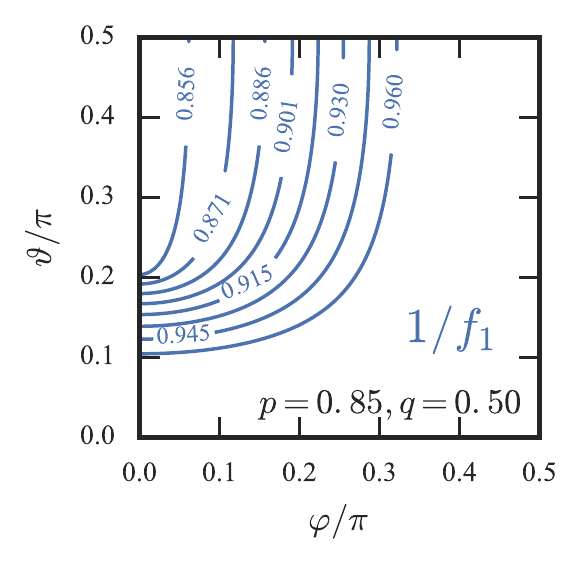}$$}

\caption{Relationship between the observed half-light radius and the
  radius used in our mass estimator formula. An ellipsoid is
  observed at spherical polar angles $(\vartheta,\varphi)$ with
  respect to its intrinsic Cartesian coordinates $(x,y,z)$ aligned
  with the principal axes. The resulting projection is an ellipse
  (shown below) with major-axis length $R_\mathrm{h}$ which lies in
  the $(x',y')$ plane of the observed Cartesian coordinate system
  $(x',y',z')$. Above the ellipsoid, we show the sphere with the
  equivalent volume as the ellipsoid. The major-axis of the ellipse is
  related to the major-axis of the ellipsoid by the factor $1/f_1$
  which is shown in the lower two panels for a prolate spheroid with
  $p=q=0.5$ (left) and an ellipsoid with axis ratios $p=0.85$ and
  $q=0.5$ (right).}
\label{Fig::Diagram}
\end{figure}

We decompose the radial scaling into two components, $f_r=f_1
f_2$. $f_2$ describes the relationship between the ellipsoidal
major-axis length and the parent spherical radius so (as described above) $f_2=(pq)^{1/3}$. The other factor $f_1$ gives the
relationship between the observed major-axis length of the half-light
ellipse $\Rh$ and the intrinsic major-axis length of the
corresponding ellipsoid $r_\mathrm{maj}$. In the spherical case, these
quantities are equal. In the ellipsoidal case, the relationship
between these quantities depends on the viewing angles and the
intrinsic shape $f_1=f_1(\vartheta,\varphi,p,q)$.  We approximate
$f_1$ by the relationship between the major-axis length of an
ellipsoid and the major-axis length of its projected ellipse. This
neglects any subtleties related to the extended nature of the true
density distribution. However, if the 3D stellar light profile falls
off sufficiently rapidly then our relationship is a good
approximation.

To derive our approximation for $f_1$, we use a coordinate
system $(x',y',z')$ related to the intrinsic coordinate system by (see Fig.
~\ref{Fig::Diagram})
\begin{equation}
\begin{split}
x&=-x'\sin\varphi-y'\cos\vartheta\cos\varphi+z'\sin\vartheta\cos\varphi\\
y&=x'\cos\varphi-y'\cos\vartheta\sin\varphi+z'\sin\vartheta\sin\varphi\\
z&=y'\sin\vartheta+z'\cos\vartheta.
\end{split}
\end{equation}
We consider the set of points where the ellipsoidal surface is
tangential to $\hat{\boldsymbol{z}}'$ which results in a rotated
ellipse in the $(x',y')$ plane. We diagonalize the resultant quadratic
surface to find the major axis length $\Rh$ as
\begin{equation}
f_1^{-2}=(\Rh/r_\mathrm{maj})^2= 2C/(A-\sqrt{B}),
\label{Eqn::Fsig_exp}
\end{equation}
where
\begin{eqnarray}
A&=&(1-q^2)\cos^2\vartheta+(1-p^2)\sin^2\vartheta\sin^2\varphi+p^2+q^2,\nonumber\\
B&=&[(1-q^2)\cos^2\vartheta-(1-p^2)\sin^2\vartheta\sin^2\varphi-p^2+q^2]^2\nonumber\\
&+& 4(1-p^2)(1-q^2)\sin^2\vartheta\cos^2\vartheta\sin^2\varphi,\nonumber\\
C&=& p^2\cos^2\vartheta+q^2\sin^2\vartheta(p^2\cos^2\varphi+\sin^2\varphi).
\end{eqnarray}
As given in \cite{Weijmans2014}, the observed ellipticity $\epsilon$
satisfies $(1-\epsilon)^2=(A-\sqrt{B})/(A+\sqrt{B})$.

For an oblate spheroid ($p=1$), eqn~(\ref{Eqn::Fsig_exp}) simplifies to
$\Rh=r_\mathrm{maj}$. For a prolate spheroid $p=q$, so we find
\begin{equation}
f_1^{-2}=\cos^2\vartheta+\sin^2\vartheta(q^2\cos^2\varphi+\sin^2\varphi).
\end{equation}
In Fig~\ref{Fig::Diagram}, we show the major axis length for a
prolate figure and a triaxial figure as a function of the viewing
angle.

The ellipsoidal half-light radius $m_h$ is well approximated by
$\tfrac{4}{3}f_r\Rh$ which should be compared to the radius of
$\tfrac{4}{3}\Rh\sqrt{1-\epsilon}$ that is empirically used \citep[e.g.][]{Koposov2015,Sanders2016} as $\Rh\sqrt{1-\epsilon}$ approximately reproduces the circularly-averaged
half-light radius of the dSph\footnote{For example, a flattened
($q=1-\epsilon$) Plummer surface profile produces a circularly-averaged
half-light radius equal to $\Rh\sqrt{\tfrac{1}{6}(1+q^2+\sqrt{1+14q^2+q^4})}$
which for small flattenings is
$\Rh(1-\tfrac{1}{2}\epsilon+\mathcal{O}(\epsilon^4))$ so well approximated by
$\Rh\sqrt{1-\epsilon}$.}.

\subsection{Near-spherical limits}\label{Sect::NearSph}

Using eqn~(\ref{Eqn::Fsig_exp}), we can find the modification factor
$f_\sigma f_r$ for the simple cases of viewing down the principal axes
of a near-spherical triaxial ellipsoid and compare to the alternative
factor $\sqrt{1-\epsilon}$. When viewing down the major axis
($\vartheta=\pi/2,\varphi=0$), we find
\begin{equation}
f_\sigma f_r\approx1+\tfrac{2}{5}(1-p)-\tfrac{3}{5}(1-q).
\end{equation}
The observed ellipticity $\epsilon=1-p/q$ so the \emph{circularized}
factor $\sqrt{1-\epsilon}\approx1+\tfrac{1}{2}(1-p)-\tfrac{1}{2}(1-q)$ which
is a close approximation to our factor $f_\sigma f_r$. Similarly, for
viewing down the intermediate axis, we find
\begin{equation}
f_\sigma f_r\approx1+\tfrac{1}{5}(1-p)-\tfrac{3}{5}(1-q),
\end{equation}
whilst $\sqrt{1-\epsilon}=\sqrt{q}\approx1-\tfrac{1}{2}(1-q)$. Finally,
viewing down the minor axis, we find
\begin{equation}
f_\sigma f_r\approx1-\tfrac{3}{5}(1-p)+\tfrac{1}{5}(1-q),
\end{equation}
whilst $\sqrt{1-\epsilon}=\sqrt{p}\approx1-\tfrac{1}{2}(1-p)$. We note that
the flattening in the line-of-sight direction (e.g. $p$ in the
intermediate axis case) has a smaller contribution to the factor
$f_\sigma f_r$. This demonstrates that the simple factor
$\sqrt{1-\epsilon}$ goes a long way to account for the velocity and
radial scalings we propose.

\subsection{Results}

\begin{figure}
$$\includegraphics[width=0.9\columnwidth]{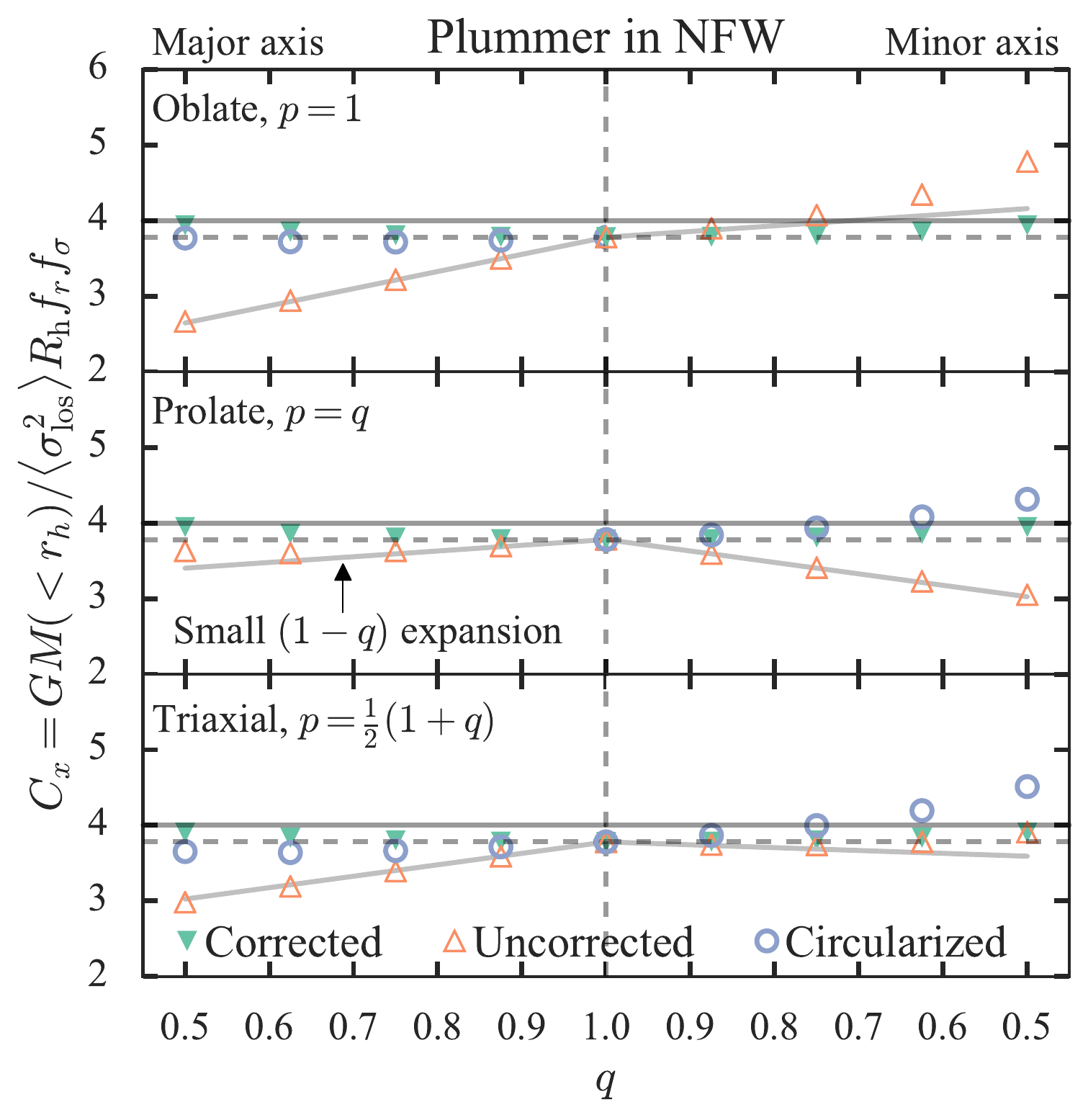}$$
\caption{Mass estimator constant for the half-light ellipsoid against the
flattening $q$ for Plummer models embedded in equivalently flattened NFW halos.
The top panel shows an oblate model, the middle panel a prolate
model and the bottom panel a triaxial model with $p=\tfrac{1}{2}(1+q)$. The
left half of the plot corresponds to viewing down the major axis whilst the
right half corresponds to viewing down the minor axis. The \emph{corrected} green filled
triangles show the constant from the mass estimator formula given in this
paper, the \emph{uncorrected} orange empty triangles show the constant using
the spherical mass estimator (i.e. $f_r f_\sigma=1$) and the purple circles
show the constant using the spherical mass estimator with the
\emph{circularized} radius $\Rh\sqrt{1-\epsilon}$ (not shown in the panels
where $\epsilon=0$). The horizontal solid line shows the
\protect\citeauthor{Wolf2010} constant and the dashed line shows the
constant from the spherical model for this exact case. The other gray solid
lines show the small $(1-q)$ expansion of $C_x f_\sigma f_r$ (i.e. the
uncorrected constant).}
\label{Fig::CorrNoCorr}
\end{figure}

In Fig.~\ref{Fig::PlummerNFW_rrat1}, we show the mass estimates using our
formulae for an oblate and prolate model viewed edge-on. The models have the
same ellipsoidal mass profile as the spherical model shown. The factors we
have introduced correctly deproject the observed quantities producing an
unbiased mass estimate. Fig.~\ref{Fig::CorrNoCorr} shows the constant in the
half-light ellipsoid mass estimator (eqn.~\eqref{Eqn::WalkerWolf_Flattened})
for three models of flattened Plummer profiles embedded in equivalently
flattened NFW halos with $m_\mathrm{DM}/m_\star=5$. We show an oblate, prolate
and triaxial $p=\tfrac{1}{2}(1+q)$ model. Simply using the spherical mass
estimator with $R_\mathrm{h}$ underestimates/overestimates the ellipsoidal
mass for the oblate/prolate case viewed face-on (down the minor/major axis).
Similarly, the edge-on case (major for oblate, minor for prolate) produces
overestimates of the mass for both oblate and prolate models. For the triaxial
model the spherical mass estimator produces an overestimate when viewing down
the major axis and (for this particular case) is largely unbiased when viewing
down the minor axis. The results using the correction factors $f_\sigma$
and $f_r$ are unbiased estimates of the mass within the ellipsoid
$m=\tfrac{4}{3}m_\star$ and using the spherical mass estimator with the
`circularized' radius $R_\mathrm{h}\sqrt{1-\epsilon}$ produces very similar
results to the corrected version. This echoes a result in \cite{Sanders2016}
who demonstrated that the correction to the D-factor (important for
interpreting dark-matter decay signals) is almost independent of the
flattening for edge-on systems. The near-spherical expansions of
\S~\ref{Sect::NearSph} are also shown, which replicate the trends over
the full $q$ range.

\begin{figure}
$$\includegraphics[width=\columnwidth]{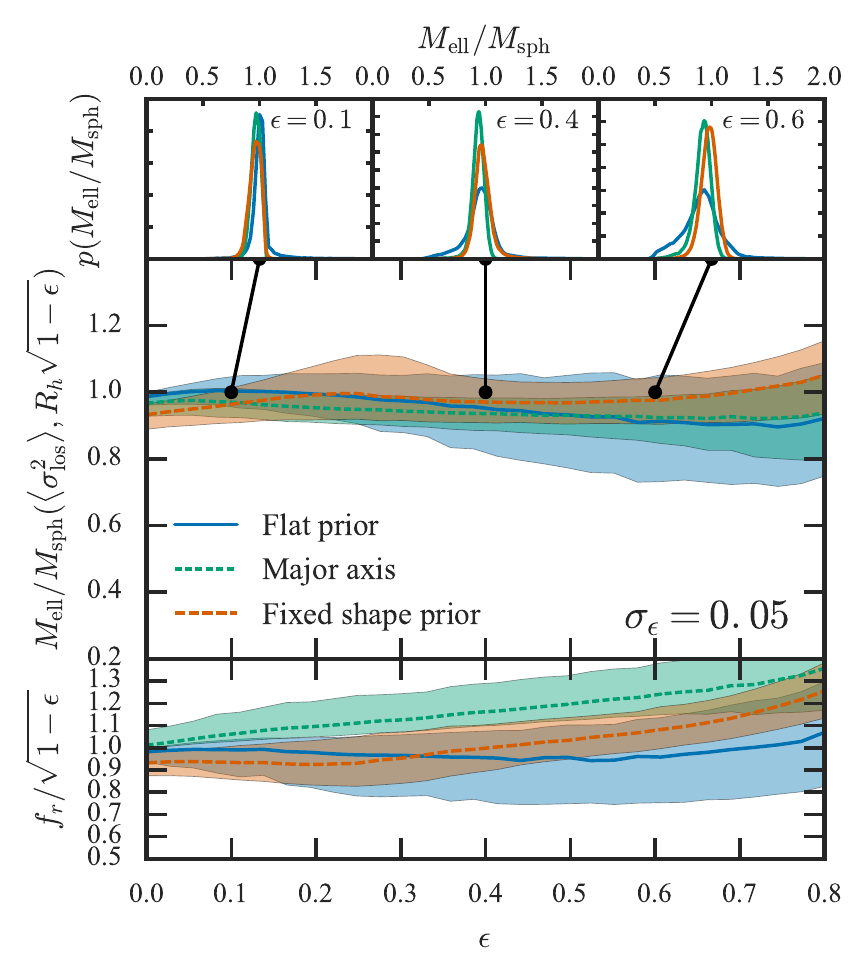}$$
\caption{Ratio of the mass within the half-light ellipsoid to the mass
  estimated from the spherical mass estimator
  \protect\citep{Walker2009,Wolf2010} using the \emph{circularized}
  half-light radius $R_\mathrm{h}\sqrt{1-\epsilon}$ against
  ellipticity. At each ellipticity we show the median and $\pm1\sigma$
  spreads in the mass estimates for triaxial figures within a $0.05$
  spread of the required ellipticity and with fixed $R_\mathrm{h}$ and
  $\slosT^{1/2}$ under three prior assumptions: the blue solid line
  corresponds to a uniform prior on viewing angle, intrinsic
  ellipticity and triaxiality, the green short-dashed line corresponds
  to a prior of preferentially viewing down the major-axis and the
  orange long-dashed line corresponds to a fixed triaxiality and
  intrinsic ellipticity prior. The three histograms show slices
  through the main figure at $\epsilon=0.1, 0.4$ and $0.6$. The lower
  panel shows the median and $\pm1\sigma$ brackets of $f_r/\sqrt{1-\epsilon}$
  -- the ratio of the dSph radius to the `circularized' radius. }
\label{Fig::massaverage}
\end{figure}

Our proposed modifications correctly reproduce the mass within
ellipsoids. However, this relies on knowing the intrinsic shape and
alignment of the dSph. Such information is not accessible, but we can
put priors on possible models which reproduce the observables. We
choose to put priors on the triaxiality $T=(1-p^2)/(1-q^2)$,
flattening $q$ and the viewing angles $(\vartheta,\varphi)$. We
consider three priors:
\begin{enumerate}
\item Flat prior -- $T\sim\mathcal{U}(0,1)$, $q\sim\mathcal{U}(0.05,1)$, $\cos\vartheta\sim\mathcal{U}(0,1)$, $\varphi\sim\mathcal{U}(0,\pi/2)$.
\item Major-axis prior -- $T\sim\mathcal{U}(0,1)$, $q\sim\mathcal{U}(0.05,1)$, $\vartheta\sim\mathcal{N}(\pi/2,0.1\mathrm{\,rad})$, $\varphi\sim\mathcal{U}(0,0.1\mathrm{\,rad})$.
\item Fixed-shape prior -- $T\sim\mathcal{N}(0.55,0.04)$, $q\sim\mathcal{N}(0.49,0.12)$, $\cos\vartheta\sim\mathcal{U}(0,1)$, $\varphi\sim\mathcal{U}(0,\pi/2)$.
\end{enumerate}
where the final prior is taken from a fit to the shapes of the Local
Group dSphs from \cite{SanchezJanssen2016}. The major-axis prior is
inspired by the observation from simulations that the major-axes of
subhaloes points towards the centre of the host halo
\cite[e.g.][]{Barber2015}.  We sample from the priors folded with a
normal distribution on the observed ellipticity with width
$\sigma_\epsilon=0.05$ \citep[using \emph{emcee}][]{ForemanMackey2012}
and for each sample compute the mass within the half-light ellipsoid
from equation~\eqref{Eqn::WalkerWolf_Flattened}. The results for a range
of observed ellipticities are shown in Fig.~\ref{Fig::massaverage}. We
show the mass estimates over the spherical mass estimator using the
`circularized' radius. We see that using the spherical mass estimator
in this way reproduces the mass within the half-light ellipsoid over
the full range of ellipticities\footnote{We note that a similar observation
was made by \cite{Laporte2013} who found that the variations of $\Rh$ and
$\slosT$ with triaxiality compensated each other to give an unbiased mass
estimate.}. The uncertainty in the estimator
increases with increasing ellipticity but is only $\sim10-20\percent$
for $\epsilon\sim0.4$ (a typical dSph flattening). There is the
tendency for the mass within the half-light ellipsoid to be
overestimated for large $\epsilon$, but only by $\sim5\percent$.
We also show the distribution of $f_r/\sqrt{1-\epsilon}$ for each
prior assumption (i.e. the ratio of the size of the ellipsoid to the
$\Rh\sqrt{1-\epsilon}$ approximation). For the uniform prior,
this ratio is unity (within $\sim10-20\percent$) so the `size' of the dSphs
are well approximated by $\Rh\sqrt{1-\epsilon}$. For the other two priors, the
ratio increases with ellipticity as the intrinsic ellipsoids are on average
more elongated along the line-of-sight so larger than $\Rh\sqrt{1-\epsilon}$.

We have demonstrated that the mass within the half-light ellipsoid can
be accurately estimated using the spherical mass estimator
formulae. Although we do not know the shape or
orientation of this half-light ellipsoid, we can say with confidence
the mass within it. Therefore, we can accurately estimate the
mass-to-light ratio using the mass within the half-light ellipsoid and
half the total luminosity $L$. {\it We conclude that using the
spherical mass estimators \citep{Walker2009,Wolf2010} with the
`circularized' half-light radius produces accurate estimates of the
mass-to-light ratio of dSphs, irrespective of flattening, provided the light and dark matter are stratified on the same self-similar concentric ellipsoids}.

\section{Conclusions}

This {\it Letter} has answered the question: how should the mass of a
flattened, dispersion-supported galaxy like a dwarf spheroidal be
estimated? If the galaxy were spherical, then the answer is
well-established.  Accurate mass estimators depending on the observable
half-light radius and the velocity dispersion of the stars have been
devised by a number of investigators
\citep{Walker2009,Wolf2010,Amorisco2012,Campbell2016}.

We have shown how to modify the spherical mass estimators so that they
work for flattened systems in which the light and dark
matter are stratified on the same concentric self-similar ellipsoids. This represents a limiting case as simulations indicate the dark matter
distribution is in fact rounder than the light \citep{Ab10,Ze12} due to
baryonic feedback effects, particularly for the more massive dSphs. The
modifications require knowledge of the intrinsic shape and alignment
of the triaxial figure and reproduce the mass within ellipsoids by
deprojecting the half-light radius and line-of-sight velocity
dispersion. The resulting mass estimates are independent of details of
the radial profile and are as accurate as the corresponding spherical
formulae.

This would be of little use if we require knowledge of intrinsic
properties. However, we have also shown that, when averaging over
triaxial configurations that are consistent with the observed
ellipticity $\epsilon$, major-axis half-light length $R_\mathrm{h}$
and line-of-sight velocity dispersion, the mass within the half-light
ellipsoid is well approximated by the spherical mass estimate using
the `circularized' half-light radius of
$R_\mathrm{h}\sqrt{1-\epsilon}$. The scatter in the estimate increases
with ellipticity but is only $10-20\percent$ for $\epsilon\sim0.6$. In
turn, this observation implies that mass-to-light ratios using
spherical estimators, together with a luminosity of $L_{1/2}=L/2$, are
accurate and insensitive to the flattening of the dSph. This therefore
provides a surprisingly simple, flexible and effective way to account
for the effects of flattening.

\label{lastpage}

\begin{thebibliography}{}
\expandafter\ifx\csname natexlab\endcsname\relax\def\natexlab#1{#1}\fi

\bibitem[{{Abadi} {et~al.}(2010){Abadi}, {Navarro}, {Fardal}, {Babul}, \&
  {Steinmetz}}]{Ab10}
{Abadi}, M.~G., {Navarro}, J.~F., {Fardal}, M., {Babul}, A., \& {Steinmetz}, M.
  2010, \mnras, 407, 435

\bibitem[{{Amorisco} \& {Evans}(2012)}]{Amorisco2012}
{Amorisco}, N.~C., \& {Evans}, N.~W. 2012, \mnras, 419, 184

\bibitem[{{Barber} {et~al.}(2015){Barber}, {Starkenburg}, {Navarro}, \&
  {McConnachie}}]{Barber2015}
{Barber}, C., {Starkenburg}, E., {Navarro}, J.~F., \& {McConnachie}, A.~W.
  2015, \mnras, 447, 1112

\bibitem[{{Binney} \& {Tremaine}(2008)}]{BinneyTremaine}
{Binney}, J., \& {Tremaine}, S. 2008, {Galactic Dynamics: Second Edition}
  (Princeton University Press)

\bibitem[{{Campbell} {et~al.}(2016){Campbell}, {Frenk}, {Jenkins}, {Eke},
  {Navarro}, {Sawala}, {Schaller}, {Fattahi}, {Oman}, \&
  {Theuns}}]{Campbell2016}
{Campbell}, D.~J.~R., {Frenk}, C.~S., {Jenkins}, A., {et~al.} 2016, ArXiv
  e-prints, arXiv:1603.04443

\bibitem[{{Foreman-Mackey} {et~al.}(2013){Foreman-Mackey}, {Conley},
  {Meierjurgen Farr}, {Hogg}, {Long}, {Marshall}, {Price-Whelan}, {Sanders}, \&
  {Zuntz}}]{ForemanMackey2012}
{Foreman-Mackey}, D., {Conley}, A., {Meierjurgen Farr}, W., {et~al.} 2013,
  {emcee: The MCMC Hammer}, , , astrophysics Source Code Library, ascl:1303.002

\bibitem[{{Koposov} {et~al.}(2015){Koposov}, {Belokurov}, {Torrealba}, \&
  {Evans}}]{Koposov2015}
{Koposov}, S.~E., {Belokurov}, V., {Torrealba}, G., \& {Evans}, N.~W. 2015,
  \apj, 805, 130

\bibitem[{{Laporte} {et~al.}(2013){Laporte}, {Walker}, \&
  {Pe{\~n}arrubia}}]{Laporte2013}
{Laporte}, C.~F.~P., {Walker}, M.~G., \& {Pe{\~n}arrubia}, J. 2013, \mnras,
  433, L54

\bibitem[{{Roberts}(1962)}]{Roberts1962}
{Roberts}, P.~H. 1962, \apj, 136, 1108

\bibitem[{{S{\'a}nchez-Janssen} {et~al.}(2016){S{\'a}nchez-Janssen},
  {Ferrarese}, {MacArthur}, {C{\^o}t{\'e}}, {Blakeslee}, {Cuillandre}, {Duc},
  {Durrell}, {Gwyn}, {McConnacchie}, {Boselli}, {Courteau}, {Emsellem}, {Mei},
  {Peng}, {Puzia}, {Roediger}, {Simard}, {Boyer}, \&
  {Santos}}]{SanchezJanssen2016}
{S{\'a}nchez-Janssen}, R., {Ferrarese}, L., {MacArthur}, L.~A., {et~al.} 2016,
  \apj, 820, 69

\bibitem[{{Sanders} {et~al.}(2016){Sanders}, {Evans}, {Geringer-Sameth}, \&
  {Dehnen}}]{Sanders2016}
{Sanders}, J.~L., {Evans}, N.~W., {Geringer-Sameth}, A., \& {Dehnen}, W. 2016,
  ArXiv e-prints, arXiv:1604.05493

\bibitem[{{Walker} {et~al.}(2009){Walker}, {Mateo}, {Olszewski},
  {Pe{\~n}arrubia}, {Wyn Evans}, \& {Gilmore}}]{Walker2009}
{Walker}, M.~G., {Mateo}, M., {Olszewski}, E.~W., {et~al.} 2009, \apj, 704,
  1274

\bibitem[{{Weijmans} {et~al.}(2014){Weijmans}, {de Zeeuw}, {Emsellem},
  {Krajnovi{\'c}}, {Lablanche}, {Alatalo}, {Blitz}, {Bois}, {Bournaud},
  {Bureau}, {Cappellari}, {Crocker}, {Davies}, {Davis}, {Duc}, {Khochfar},
  {Kuntschner}, {McDermid}, {Morganti}, {Naab}, {Oosterloo}, {Sarzi}, {Scott},
  {Serra}, {Verdoes Kleijn}, \& {Young}}]{Weijmans2014}
{Weijmans}, A.-M., {de Zeeuw}, P.~T., {Emsellem}, E., {et~al.} 2014, \mnras,
  444, 3340

\bibitem[{{Wolf} {et~al.}(2010){Wolf}, {Martinez}, {Bullock}, {Kaplinghat},
  {Geha}, {Mu{\~n}oz}, {Simon}, \& {Avedo}}]{Wolf2010}
{Wolf}, J., {Martinez}, G.~D., {Bullock}, J.~S., {et~al.} 2010, \mnras, 406,
  1220

\bibitem[{{Zemp} {et~al.}(2012){Zemp}, {Gnedin}, {Gnedin}, \&
  {Kravtsov}}]{Ze12}
{Zemp}, M., {Gnedin}, O.~Y., {Gnedin}, N.~Y., \& {Kravtsov}, A.~V. 2012, \apj,
  748, 54

\end{thebibliography}


\clearpage
\end{document}